\documentclass[review,12pt]{elsarticle}

\usepackage{amssymb}
\usepackage{amsthm}
\usepackage{amsmath}
\usepackage{mathrsfs}
\usepackage{graphicx}
\usepackage{epstopdf}
\usepackage{float}
\usepackage{caption}
\usepackage{subcaption}
\usepackage{bm}
\usepackage{bbm}
\usepackage{mathrsfs}
\usepackage{cleveref}
\usepackage{soul}
\usepackage{color,soul} 
\usepackage{color} 
\usepackage{siunitx}
\usepackage[margin=2.5cm]{geometry}
\biboptions{sort&compress}

\journal{ }

\makeatletter
\def\@author#1{\g@addto@macro\elsauthors{\normalsize%
    \def\baselinestretch{1}%
    \upshape\authorsep#1\unskip\textsuperscript{%
      \ifx\@fnmark\@empty\else\unskip\sep\@fnmark\let\sep=,\fi
      \ifx\@corref\@empty\else\unskip\sep\@corref\let\sep=,\fi
      }%
    \def\authorsep{\unskip,\space}%
    \global\let\@fnmark\@empty
    \global\let\@corref\@empty  
    \global\let\sep\@empty}%
    \@eadauthor={#1}
}
\makeatother

\begin{document}

\begin{frontmatter}



\title{Phase field fracture predictions of microscopic bridging behaviour of composite materials}


\author{Wei Tan\fnref{QMUL}}

\author{Emilio Mart\'{\i}nez-Pa\~neda\corref{cor1}\fnref{IC}}
\ead{e.martinez-paneda@imperial.ac.uk}

\address[QMUL]{School of Engineering and Materials Science, Queen Mary University London, Mile End Road, London, E1 4NS, UK}

\address[IC]{Department of Civil and Environmental Engineering, Imperial College London, London SW7 2AZ, UK}

\cortext[cor1]{Corresponding author.}

\begin{abstract}
We investigate the role of microstructural bridging on the fracture toughness of composite materials. To achieve this, a new computational framework is presented that integrates phase field fracture and cohesive zone models to simulate fibre breakage, matrix cracking and fibre-matrix debonding. The composite microstructure is represented by an embedded cell at the vicinity of the crack tip, whilst the rest of the sample is modelled as an anisotropic elastic solid. The model is first validated against experimental data of transverse matrix cracking from single-notched three-point bending tests. Then, the model is extended to predict the influence of grain bridging, brick-and-mortar microstructure and 3D fibre bridging on crack growth resistance. The results show that these microstructures are very efficient in enhancing the fracture toughness via fibre-matrix debonding, fibre breakage and crack deflection. In particular, the 3D fibre bridging effect can increase the energy dissipated at failure by more than three orders of magnitude, relative to that of the bulk matrix; well in excess of the predictions obtained from the rule of mixtures. These results shed light on microscopic bridging mechanisms and provide a virtual tool for developing high fracture toughness composites.

\end{abstract}


\end{frontmatter}


\section{Introduction}
\label{Introduction}

Fibre-reinforced composites are ubiquitous in many biological systems and modern engineering structures. For example, articular cartilage is a composite with a proteoglycan gel matrix reinforced by collagen fibrils, and wood is made from long cellulose fibres bonded by a much weaker matrix called lignin. These biological materials are typically tough and resilient to fracture. Inspired by nature, fibre reinforcement is widely used to improve the fracture toughness of many engineering materials such as ceramics, metals, polymers, and cementitious materials. In these reinforced composites, fibres bridge the gap between two adjacent crack surfaces and delay crack tip opening. A significant amount of fracture energy is absorbed by fibre-matrix debonding, fibre pull-out, and fibre breakage, thereby notably increasing the material fracture toughness \cite{Laffan2012,Marin2016,Tan2018}.  

%

\begin{figure}[h]
    \centering
    \includegraphics[width=1.0\textwidth]{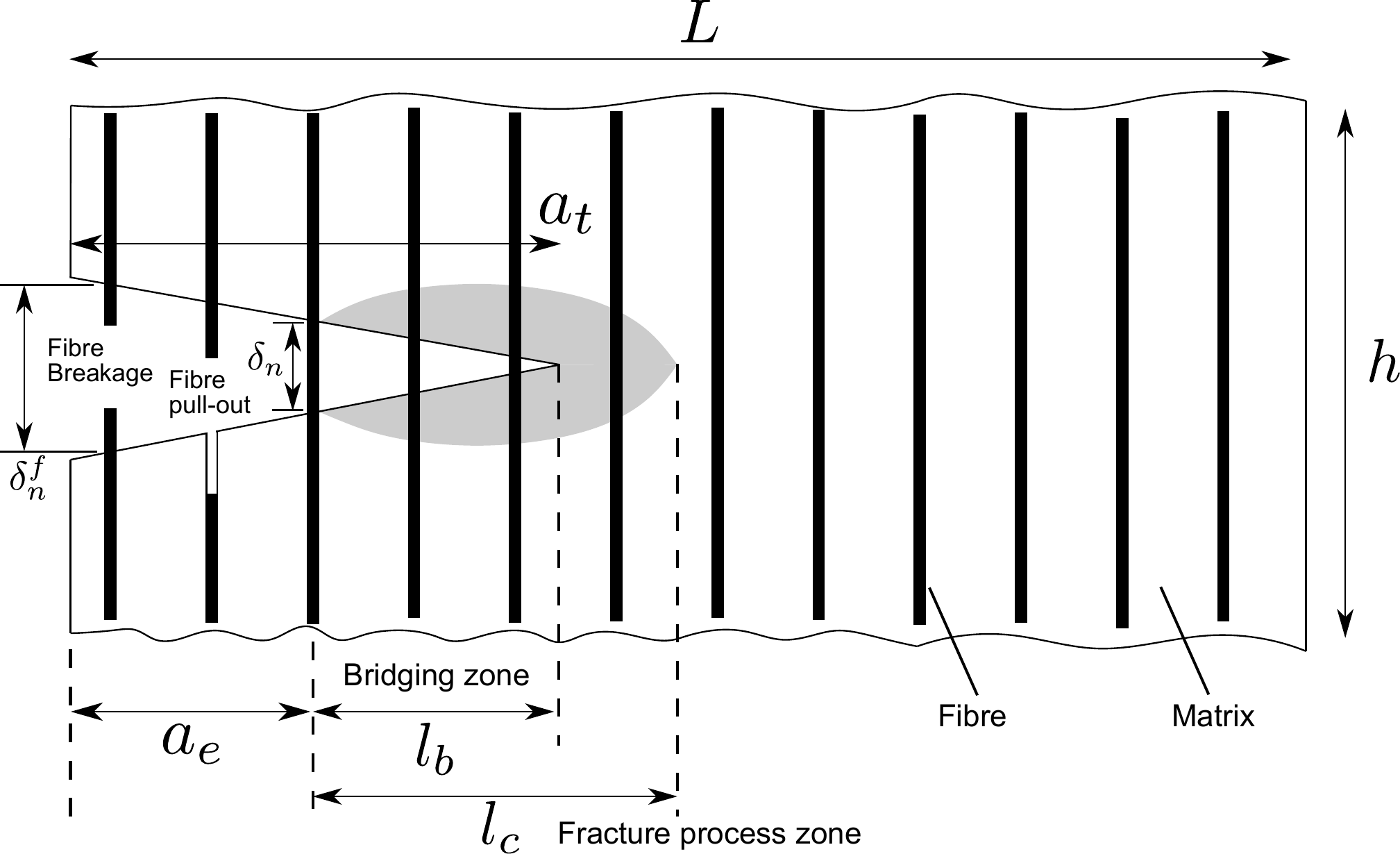}
    \caption{Sketch of the fibre bridging mechanism and the bridging zone length.}
    \label{Fig:Bridging}
\end{figure}

A sketch of the fibre bridging zone is shown in Fig. \ref{Fig:Bridging}. The fibres can transfer mechanical load and thus impede the opening of the crack. If the bridging zone length $l_b$ is much smaller than the dimensions of the sample ($L$ and $h$, see Fig. \ref{Fig:Bridging}), small-scale bridging (SSB) conditions prevail and, as a result, the R-curve can be regarded as a material property. However, when the bridging zone length is comparable to $h$ or $L$, the conditions are those of large-scale bridging (LSB). In such a case, the R-curve can no longer be interpreted as a material property, as it depends on the specimen geometry \cite{Bao1995,Sorensen1998a}. 

In small-scale bridging experiments ($l_b \ll L$), the unloading compliance $C=\delta/p$ is used to calculate the effective crack size $a_e$, following the ASTM standard \cite{ASTM1820}. This effective crack length is then used to calculate the geometrical correction factor $f(a_e/W)$ and the stress intensity factor as $K=P S(BW^{3/2})^{-1}f(a_e/W)$, where $P$ is the applied force, and $B$ and $W$ are the sample thickness and width, respectively. Finally, the $J$-integral is calculated by substituting $K$ into the plain strain equation below,
\begin{equation}\label{eq:ssbJ}
	J=\frac{K^2(1-\nu^2)}{E} \, .
\end{equation}

For large-scale bridging problems ($l_b \approx L$), the $J$ integral is evaluated locally around the fracture process zone as follows \cite{Rice1968,Bao1995,Sorensen1998a}:
\begin{equation}\label{eq:Bridging_law}
    J=J_0+\int_{0}^{\delta_n^f}\sigma_n(\delta_n)\text{d}\delta_n ,
\end{equation} 
where $\sigma_{n}$ and $\delta_n$ denote the normal stress and the opening displacement of the bridging ligament respectively.  Also, $J$ and $J_0$ denote the global and local (crack tip) energy release rates, respectively.





To date, most research efforts have used experimental techniques to investigate crack bridging effects on fracture toughness. Several crack bridging paradigms have been explored, including the use of toughened thermoplastic particles \cite{Kinloch1994}, carbon nanotubes \cite{Ou2019}, Z-pin \cite{Ravindran2019a,m2019effective} and 3D woven composite architectures  \cite{Meza2019}. However, these trial-and-error approaches are expensive and time-consuming, often leading to sub-optimal designs. Alternatively, the discovery of new architectural designs can be accelerated by using finite element-based computational micromechanics. This emerging approach resolves the micromechanical behaviour of composites by using a representative volume element (RVE) or an embedded cell of composite microstructures. The predicted homogenised behaviour, including constitutive material constants (e.g. Young's modulus, strength), is then transferred to mesoscale or macroscale models. This enables a rapid optimisation process through a direct change of the constituent (fibre, matrix and interfaces) properties \cite{Llorca2011,Tan2018,Herraez2018}.


Several numerical methods are being used in these computational micromechanics endeavours, including Continuum Damage Mechanics (CDM) models \cite{Chaboche1988,Tan2015,tan2020physically,Aoki2021}, the Extended Finite Element Method (X-FEM) \cite{Hachi2020,vellwock2018multiscale}, Cohesive Zone Models (CZM) \cite{Camanho2002} and the Floating Node Method \cite{Chen2014a}. Despite their effectiveness, these techniques are often limited in capturing complicated crack paths, which naturally result from crack coalescence, branching and bridging events. A promising alternative for modelling the progressive failure of materials is the Phase Field (PF) fracture model \cite{Bourdin2000,Miehe2010a,TAFM2020}, which is gaining increasing traction in the computational mechanics community \cite{Wu2020}. The popularity of phase field fracture methods arguably lies in their ability to predict complex fracture processes such as crack nucleation, intricate crack paths, branching and merging; key features when capturing the complex behaviour exhibited by cracks propagating through the microstructure. Moreover, these predictions are mesh-independent and do not require re-meshing or to explicitly track the crack discontinuity. 

In the PF fracture method, fracture takes place in agreement with the thermodynamics of fracture and Griffith's energy balance \cite{Griffith1920}; cracking initiates when the energy released by the solid exceeds the critical energy required to create new surfaces, the material fracture toughness $G_c$. The PF fracture model has been recently used to capture the fracture behaviour of composite materials at the microscale and mesoscale levels \cite{Quintanas-Corominas2018,Quintanas-Corominas2019,Quintanas-Corominas2019a,Zhang2019c,Espadas-Escalante2019,CPB2019,Bui2021,Tan2021a}. In this work, we build on the potential of the PF fracture method to answer two important questions that remain unaddressed: (i) what is the effect of fibre, matrix, and fibre-matrix interface properties on the crack bridging behaviours? and, considering crack bridging, (ii) what is the influence of microstructural shapes and distributions upon the crack growth resistance?  

In this study, an integrated PF-CZM computational model is proposed to predict the microscopic bridging behaviour of composite materials. The model is first validated with experimental data from single-edge notched beam bending testing. Then, we simulate the bridging behaviour of varying microstructures, including grain bridging, brick-and-mortar and 3D fibre bridging scenarios. The main novel contributions of this work are: (i) the pioneering use of a combined PF-CZM model to predict microscale bridging behaviour; (ii) the analysis, for the first time, of the effect of reinforcement (grain, fibre) toughness on crack growth resistance; and (iii) the characterization of the previously unknown effect on the fracture toughness of various microstructures including bridged-grain, brick-and-mortar, and 3D fibre bridging. Our results show that the modelling framework presented provides new pathways for an efficient and accurate design of damage-tolerant composite materials and structures.

\section{Numerical framework}
\label{Sec:NumModel}

\begin{figure}[h]
    \centering
    \includegraphics[width=1.0\textwidth]{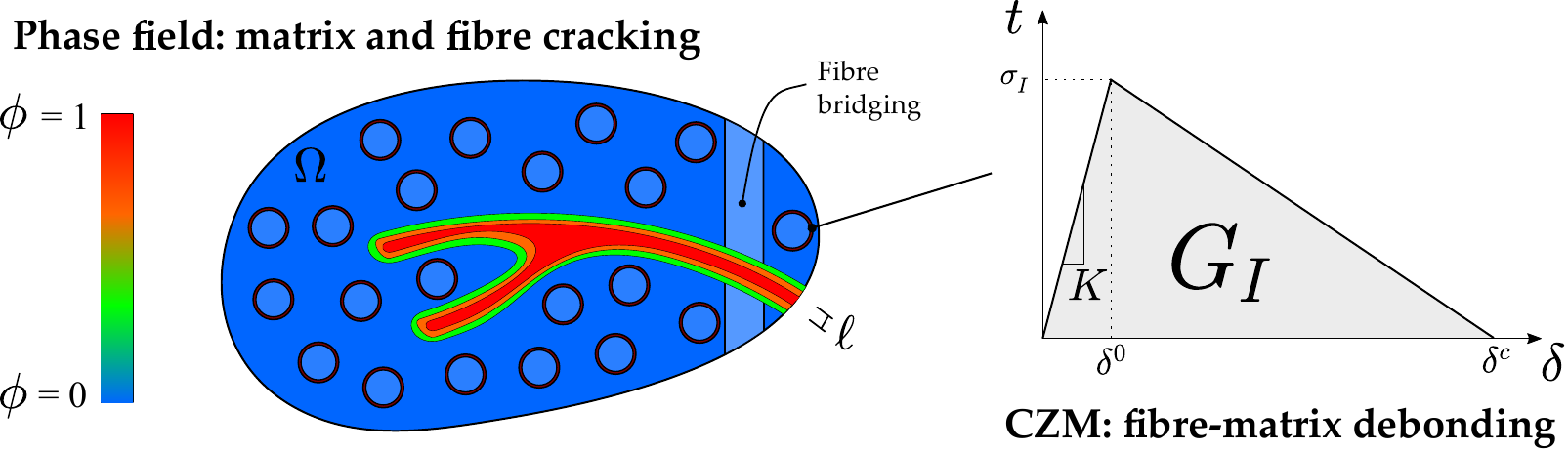}
    \caption{Sketch of the coupled phase field - cohesive zone model (CZM) framework presented. The composite microstructure is explicitly resolved, phase field fracture is used to predict matrix and fibre cracking while CZM is used to simulate fibre-matrix debonding.}
    \label{Fig:PFMCZM}
\end{figure}

Our numerical framework captures composite cracking phenomena such as fibre fracture, fibre bridging, fibre-matrix debonding and matrix cracking, in arbitrary geometries and dimensions. This is achieved by combining two fracture methodologies: (i) the phase field (PF) fracture method, and (ii) the cohesive zone model (CZM); see Fig. \ref{Fig:PFMCZM}. PF and CZM have been successfully combined to simulate 2D boundary value problems related to the failure of hybrid laminates and short fiber-reinforced composites \cite{Alessi2017,Zhang2020a,Guillen-Hernandez2020,AsurVijayaKumar2021}. Both modelling strategies are described below; no direct coupling is defined between PF and CZM, implying the assumption that matrix-fibre debonding has a negligible impact on bulk matrix and fibre cracking, and viceversa.

\subsection{Phase field fracture modelling}

The phase field fracture method builds upon the thermodynamics of fracture and the energy balance first proposed by Griffith \cite{Griffith1920,Bourdin2008}. For a crack to propagate, the reduction in potential energy resulting from crack growth has to be balanced with the increase in surface energy associated with the creation of two new free surfaces. Thus, consider a cracked solid undergoing mechanical deformation, as characterised by a strain tensor $\bm{\varepsilon}$ and an elastic strain energy density $\psi \left( \bm{\varepsilon} \right)$. Assuming for simplicity that the solid is subjected to an applied displacement, the variation of its total potential energy $\mathcal{E}$ due to an incremental increase in the crack area $\text{d}A$ must satisfy:
\begin{equation}
    \frac{\text{d}\mathcal{E}}{\text{d}A} = \frac{\text{d} \psi \left( \bm{\varepsilon} \right)}{\text{d} A} + \frac{\text{d}W_c}{\text{d}A} = 0 \, ,
\end{equation}

\noindent where $W_c$ is the work required to create two new surfaces, and its derivative with respect to the crack area is the so-called critical energy release rate $G_c=\text{d}W_c/\text{d}A$, a material property that characterises the fracture resistance of the solid. In the context of ideally brittle solids: $G_c=2 \gamma_s$, where $\gamma_s$ is the surface energy of the material. Cracking is thus dictated by the competition between the energy stored in the solid $\psi$ and the material toughness $G_c$; for a solid $\Omega$ with crack surface $\Gamma$, this energy balance can be formulated in a variational form as \cite{Francfort1998}:
\begin{equation}\label{eq:Egriffith}
    \mathcal{E} = \int_\Omega \psi \left( \bm{\varepsilon} \right) \, \text{d} V + \int_\Gamma G_c \, \text{d} \Gamma \, .
\end{equation}

Equation (\ref{eq:Egriffith}) postulates Griffith's minimality principle in a global manner. Crack growth is the result of the exchange between elastic and fracture energies, and cracking phenomena of arbitrary complexity can be captured by minimising (\ref{eq:Egriffith}). However, the minimisation of Griffith's functional $\mathcal{E}$ is hindered by the unknown nature of the crack surface $\Gamma$; characterising evolving interfaces is a well-known computational challenge. This can be overcome by the use of phase field methods, which have emerged as a promising modelling strategy to \emph{implicitly} track interfaces. The pivotal step in phase field models is the definition of an auxiliary phase field variable $\phi$, which is used to describe discrete discontinuous phenomena (such as cracks) in a diffuse fashion. The phase field $\phi$ takes two distinct values in each of the phases, e.g. 0 and 1, and varies smoothly near the interface. Such a modelling strategy has been successfully used to predict the evolution of material microstructures \cite{Provatas2011} and corrosion fronts \cite{JMPS2021}. In the case of fracture problems, the phase field can be regarded as a damage variable, going from $\phi=0$ in pristine material points to $\phi=1$ inside of the crack. A degradation function $g (\phi)=(1-\phi)^2$ is typically defined to reduce the material stiffness with evolving damage. Moreover, the phase field evolves in agreement with Griffith's energy balance. Hence, Griffith's functional $\mathcal{E}$ (\ref{eq:Egriffith}) can be approximated by the following regularised functional:
\begin{equation}
    \mathcal{E}_\ell  = \int_\Omega \left( 1 - \phi \right)^2 \psi_0 \left( \bm{\varepsilon}  \right) \, \text{d}V + \int_\Omega G_c  \left( \frac{\phi^2}{2 \ell} + \frac{\ell}{2} |\nabla \phi|^2 \right) \, \text{d}V \, ,
\end{equation}

\noindent where $\psi_0$ denotes the elastic strain energy density of the undamaged solid. The model is non-local, ensuring mesh objectivity, with a length scale parameter $\ell$ that governs the size of the fracture process zone. As rigorously proven using Gamma-convergence, the solution that constitutes a global minimum for the regularised functional $\mathcal{E}_\ell$ converges to that of $\mathcal{E}$ for a fixed $\ell \to 0^+$. Finally, the strong form can be readily derived by taking the first variation of $\mathcal{E}_\ell$ with respect to the primal kinematic variables and making use of Gauss' divergence theorem. Thus, the coupled field equations read, 
\begin{align}\label{eqn:strongForm}
\nabla \cdot \left[(1-\phi)^2  \boldsymbol{\sigma} \right] &= \boldsymbol{0}   \hspace{3mm} \rm{in}  \hspace{3mm} \Omega \nonumber \\ 
G_{c}  \left( \dfrac{\phi}{\ell}  - \ell \nabla^2 \phi \right) - 2(1-\phi) \, \psi  &= 0 \hspace{3mm} \rm{in} \hspace{3mm} \Omega  
\end{align}

\noindent The coupled system is solved using a staggered solution scheme \cite{Miehe2010a}. This modelling methodology has proven to be very useful in capturing complex cracking phenomena in composites and multi-phase materials, such as crack branching, intricate crack trajectories and merging of multiple defects \cite{TAFM2020c,Pillai2020,CMAME2021}. As shown below (Section \ref{Sec:FEMresults}), both matrix and fibre cracking can be captured.

\subsection{Cohesive zone modelling} 

Debonding between the matrix and the fibres is captured by using a cohesive zone model with a bi-linear traction-separation law. Fig. \ref{Fig:PFMCZM} depicts the traction-separation response assuming only normal tractions; in such a case, the constitutive behaviour of the cohesive zone interface is completely characterised by its shape, the initial interface modulus $K$, and two of the following parameters: the interface strength $\sigma_I$, the critical separation $\delta^c$ and the fracture energy $G_I$. However, under local mixed-mode conditions, the interaction between normal and shear tractions must also be defined. Here, we follow Camanho and Davila \cite{Camanho2002} and define an effective separation to describe the evolution of damage under a combination of normal and shear deformation:
\begin{equation}
    \delta_m = \sqrt{\langle \delta_n \rangle^2 + \delta_s^2  + \delta_t^2} \, ,
\end{equation}

\noindent where $\delta_n$ is the normal separation, and $\delta_s$ and $\delta_t$ are the two shear separation variables. The onset of damage is predicted in terms of normal ($t_n$) and shear ($t_s$, $t_t$) tractions using a quadratic nominal stress criterion:
\begin{equation}
    \left( \frac{\langle t_n \rangle}{\sigma_I^N} \right)^2 +\left( \frac{ t_s}{\sigma_I^S} \right)^2 +\left( \frac{ t_t}{\sigma_I^T} \right)^2 =1 \, ,
\end{equation}

\noindent where the associated interface strengths are given as $\sigma_I^N$, $\sigma_I^S$ and $\sigma_I^T$. Damage evolution is then defined as governed by the energetic Benzeggagh-Kenane fracture criterion, upon the assumption of the same critical fracture energy for the first and second shear directions ($G_I^S=G_I^T$). Accordingly, the mixed-mode interfacial critical energy release rate $G_C$ will be attained when,
\begin{equation}
    G_I^N+ \left( G_I^S- G_I^N \right) \left( \frac{G^S}{G^N + G^S} \right)^\eta = G_C
\end{equation}

\noindent where $\eta$ is a material parameter and $G_I^N$ is the fracture energy associated with the normal direction. Finally, we define a cohesive zone damage variable $D$ in terms of the maximum value attained by the effective displacement during the loading history $\delta_m^{max}$, and its magnitude at complete failure $\delta_m^f$ and at damage initiation $\delta_m^0$;
\begin{equation}
    D = \frac{\delta_m^f \left( \delta_m^{\text{max}}  - \delta_m^0 \right)}{\delta_m^{\text{max}} \left( \delta_m^f - \delta_m^0 \right)} \, .
\end{equation}

The cohesive zone model is implemented using a cohesive contact approach, by which the cohesive constraints are enforced at each slave node.

\section{Results}
\label{Sec:FEMresults}

The numerical framework presented in Section \ref{Sec:FEMresults} is used to shed light on the role of microscopic bridging on crack growth resistance. First, 2D benchmark simulations (see Section \ref{Sec:SENResults}) on composite plates with a single inclusion (grain) are conducted to understand the interactions of matrix cracking, interface debonding, pull-out and grain breakage. Then, 2D virtual tests of multiple inclusions are conducted to validate the model and explore the effect of various microstructures (grain-type, brick-and-mortar) under conditions of small scale bridging (SSB) - see Section \ref{Sec:SSBResults}. Lastly, in Section \ref{Sec:LSBResults}, we extend our model to 3D and simulate cracking under large scale bridging (LSB) conditions, capturing features such as fibre edge and shape effects.

\subsection{Small scale bridging analyses} 
\label{Sec:SSBResults}

\subsubsection{Single-inclusion benchmark studies} 
\label{Sec:SENResults}

\begin{figure}[H]
    \centering
    \includegraphics[width=1.0\textwidth]{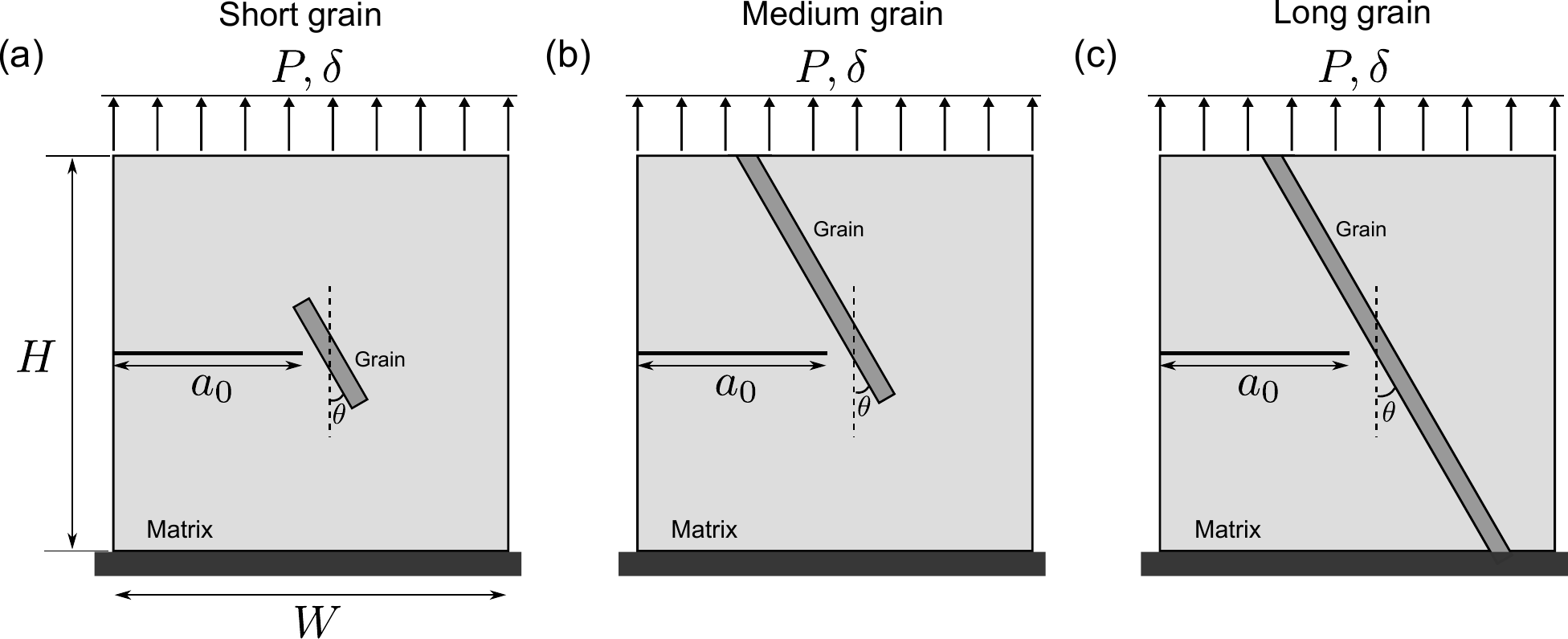}
    \caption{Single-inclusion benchmark studies: geometries, dimensions and boundary conditions of virtual tests on (a) short grain reinforced composites, (b) medium grain reinforced composites, and (c) long grain reinforced composites.}
    \label{Fig.B1}
\end{figure}

First, to gain fundamental insight into the failure behaviour of composite materials, we conduct benchmark simulations on a crack interacting with a single fibre. We model single-edge notched plates with the dimensions, geometries and boundary conditions given in Fig. \ref{Fig.B1}. The plate is made of polymer matrix reinforced by a single grain (fibre). The length and the inclination angle $\theta$ of the grain is varied. The plate is loaded by prescribing the vertical displacement in the upper side, while both vertical and horizontal displacements are fixed in the bottom edge. Phase field is used to model the fracture in the grain and bulk matrix, while the cohesive zone model is used to predict the debonding between grain and matrix. The properties of the interface are given in Table \ref{table.1}; these are used throughout this paper, unless otherwise stated. The grain has a Young's modulus of $E=74$ GPa, a Poisson's ratio of $\nu=0.35$ and a fracture toughness equal to $G_f=13.5$ J/m\textsuperscript{2}. For the matrix, we used the elastic material properties of epoxy: Young's modulus $E=$ 3.5 GPa, and Poisson's ratio $\nu=$ 0.35. The phase field length scale equals $\ell=0.005$ mm and the critical energy release rate is $G_m=10$ J/m\textsuperscript{2}.

\begin{table}[H]
    \caption{Material properties of the fibre-matrix interface \cite{Herraez2018}.} 
    \centering 
    \begin{tabular}{c c c c c c c} 
    \hline\hline 
    $\sigma_I^N$ (MPa) & $\sigma_I^S$ (MPa) & $K^N$ (GPa/mm) & $K^S$  (GPa/mm) & $G_I^N $ (J/m\textsuperscript{2}) & $G_I^S $ (J/m\textsuperscript{2}) & $\eta$\\ [0.5ex] 
    \hline 
    40 & 60 & 1000 & 1000 & 125 & 150 & 1.2\\ [1ex] 
    \hline 
    \end{tabular}
    \label{table.1} 
\end{table}

\begin{figure}[H]
    \centering
    \includegraphics[width=1.0\textwidth]{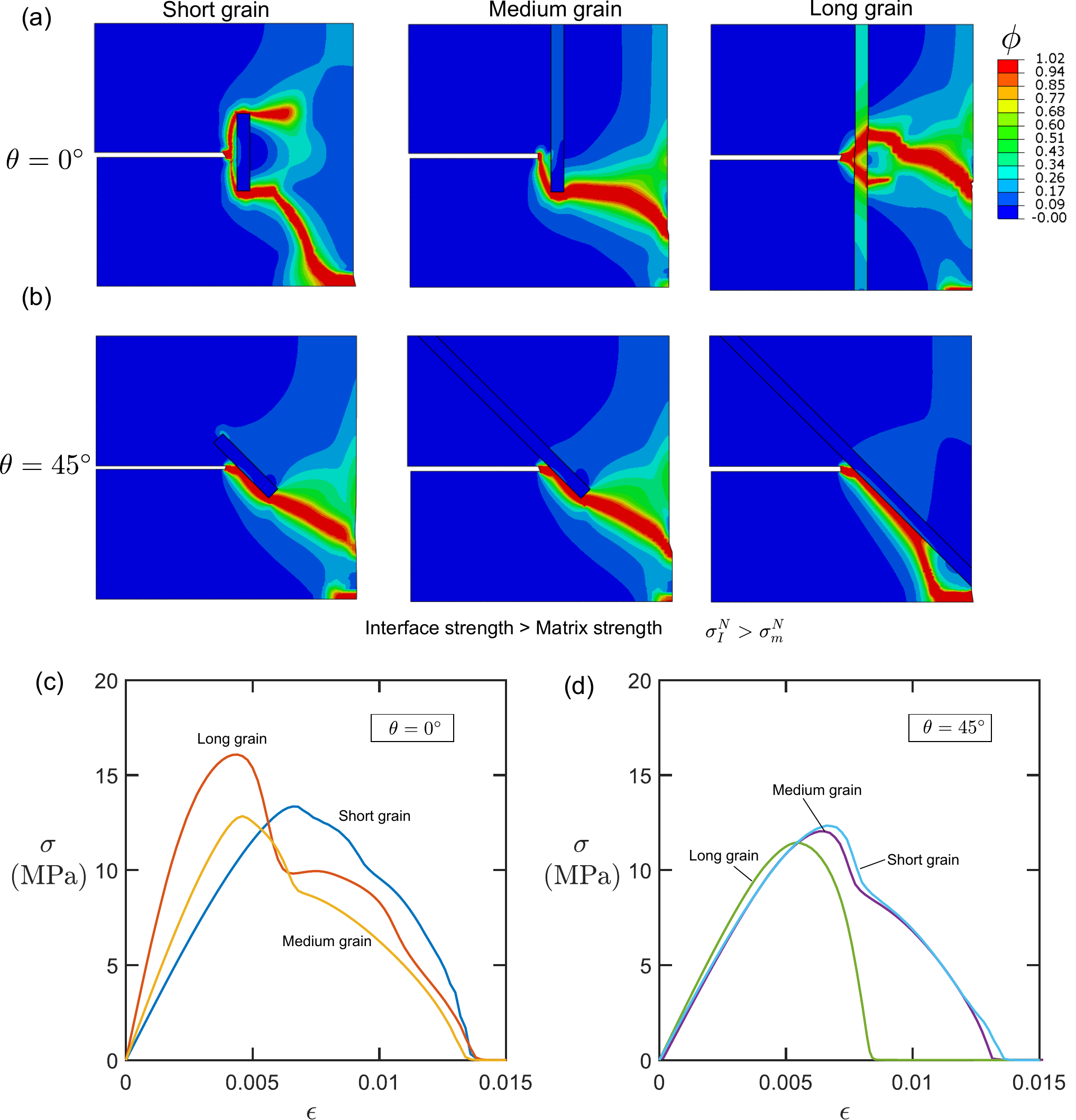}
    \caption{Single-inclusion benchmark studies. Phase field contours for fibres of varying lengths oriented (a) straight ($\theta=0^\circ$) and (b) inclined ($\theta=45^\circ$). Stress-strain responses for (c) straight grains and (d) inclined grains.}
    \label{Fig.B2}
\end{figure}

The predicted phase field fracture and stress-strain responses in Fig. \ref{Fig.B2}. The role of the grain length on crack propagation is evident in the straight grains ($\theta=0^\circ$), as shown in Fig. \ref{Fig.B2}(a). For the short and medium grains, matrix cracking initiates at the notch tip and is deflected to the grain-matrix interface. When the grain length increases to the full height of the plate, the crack goes across the fibre and branches. The influence of the fibre orientation can be observed by comparing Figs. \ref{Fig.B2}(a) and \ref{Fig.B2}(b). When the grain is inclined ($\theta=45^\circ$), cracking occurs along the grain-matrix interface, following the grain orientation. The effect of fibre length on the stress-strain responses is shown in Figs. \ref{Fig.B2}(c) and \ref{Fig.B2}(d). For the case of straight fibres ($\theta=0^\circ)$, Fig. \ref{Fig.B2}(c)), a higher stiffness, critical strength and toughness is attained for the long grain case, due to crack bridging and branching. Differences are smaller for the case of inclined fibers ($\theta=45^\circ)$, Fig. \ref{Fig.B2}(c)), with the long grain case exhibiting an earlier failure.

In addition, numerical experiments are conducted to investigate the role of the interface strength. The results (not shown) agree with expectations - if the interface strength is lowered, interface debonding is favoured and pull-out events are observed for short and medium straight grains. These pull-out events occur when interface debonding precedes matrix cracking.

\subsubsection{Experimental validation: Transverse crack} 

\begin{figure}[H]
    \centering
    \includegraphics[width=1.0\textwidth]{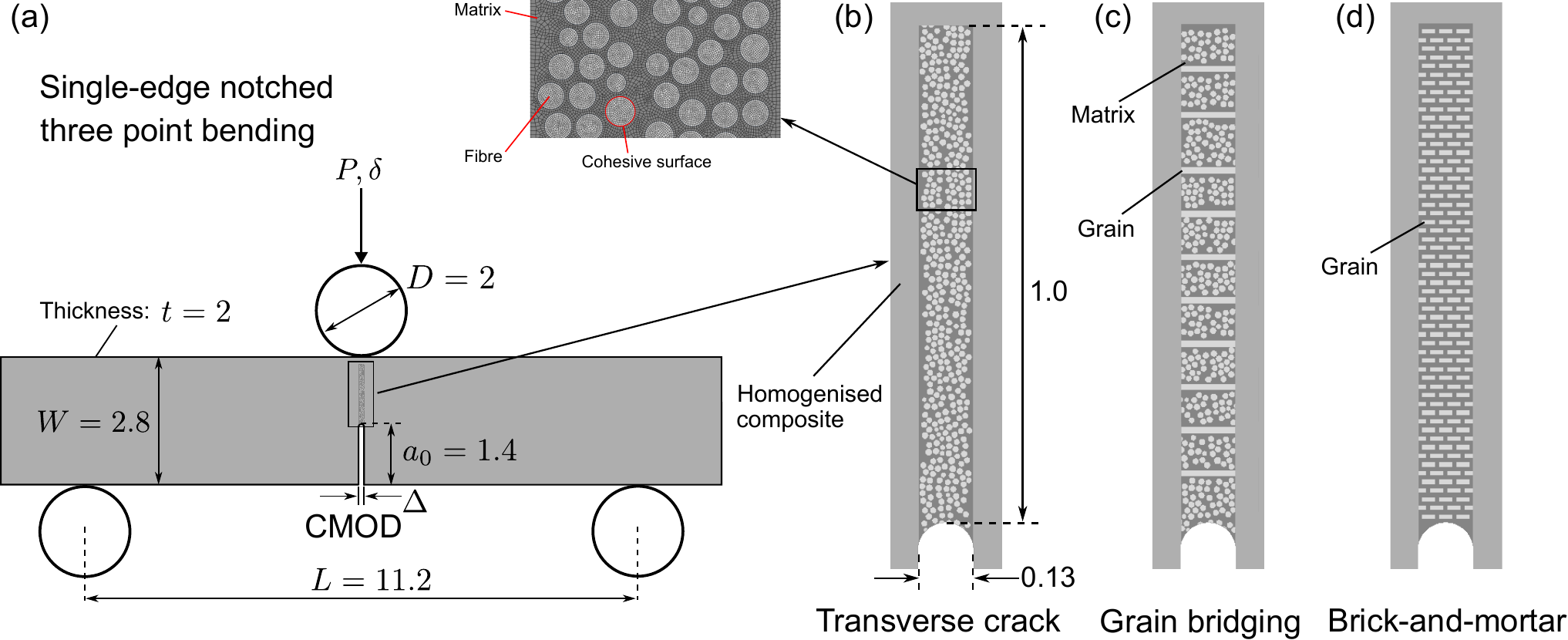}
    \caption{Single-edge notched bending tests: (a) model set-up; and microstructures of (b) fibre-reinforced polymer composites, (c) bridged-grains, and (d) brick-and-mortar. All dimensions are in mm.}
    \label{Fig.2}
\end{figure}

Now, we validate the model by conducting virtual tests of single-edge notched three-point bending (TPB) experiments and comparing them with the laboratory measurements by Canal et al. \cite{Canal2012}. To reduce the computational cost, an embedded cell method is adopted, similar to the approach developed in Ref. \cite{Herraez2018}. The composite microstructure is resolved ahead of the crack tip as an embedded cell, while the remaining ply material is homogenised as a transversely-isotropic elastic solid, see Fig. \ref{Fig.2}. A continuous displacement field between the homogenised region and the embedded cell is enforced by sharing nodes at their interface. The experimental setup and specimen geometry are given in Fig. \ref{Fig.2}. In the finite element model, the horizontal displacement is restricted in the left support to prevent rigid body motion. The initial crack length equals $a_0=$ 1.4 mm. Within the embedded cell, glass fibres of volume fraction $f_g=$ 54 \% are randomly distributed and enclosed by a polymer matrix. The fibres have a diameter that ranges from 13 $\si{\micro\metre}$ to 17 $\si{\micro\metre}$. Python scripting within ABAQUS is used to generate the various microstructures employed in this study. Specifically, the so-called random sequential adsorption (RSA) approach is used, by which a confined space is filled with random inclusions. Every time a new fibre is generated, a compatibility check is performed to determine if it overlaps with any of the existing fibres. To accelerate fibre generation and achieve high volume fractions, we use a gradient-based numerical optimisation algorithm, with the objective function being the fibre volume fraction.

The homogenised region has a Young's modulus of $E_h=$ 11 GPa and Possion's ratio equal to $\nu_h=$ 0.3. Inside the embedded zone, both glass fibre and epoxy matrix are assumed to be isotropic, elastic solids. The E-glass fibre has a Young's modulus of $E=$74 GPa, a Poisson's ratio of $\nu=$ 0.35 and a fracture toughness equal to $G_f=$ 13.5 J/m\textsuperscript{2}. For the case of the epoxy, the elastic material properties read: Young's modulus $E=$ 3.5 GPa, and Poisson's ratio $\nu=$ 0.35. The evolution of phase field damage is characterised by a phase field length scale of $\ell=0.001$ mm and critical energy release rate $G_m=$ 10 J/m\textsuperscript{2}; these values provide the best agreement with experimental observations, with $\ell$ playing only a secondary role as the process is mainly toughness-dominated due to the initial pre-crack \cite{PTRSA2021}. A cohesive surface contact between fibre and matrix is defined using a quadratic traction-separation law. The properties of the interface are from the Table \ref{table.1}. The characteristic size of the smallest element in the embedded region equals 1  $\si{\micro\metre}$ and gradually increases to 0.2 mm at the interface between the embedded cell and the homogenised solid. In total, approximately 150,000 four-node quadrilateral plane strain elements are used in each of the three microstructure case studies. 

During the virtual tests, we record the load, the crack tip opening displacement (CMOD), and the crack tip position. The current crack tip position is defined as the location of the damaged element ($\phi \geq 0.95$) that is located farthest relative to the initial crack tip (i.e., the damaged element with the highest vertical coordinate). The results obtained are shown below for the case of the experimental validation and the analysis of grain bridging and brick-and-mortar microstructures. 

\begin{figure}[htbp]
    \centering
    \includegraphics[width=1.0\textwidth]{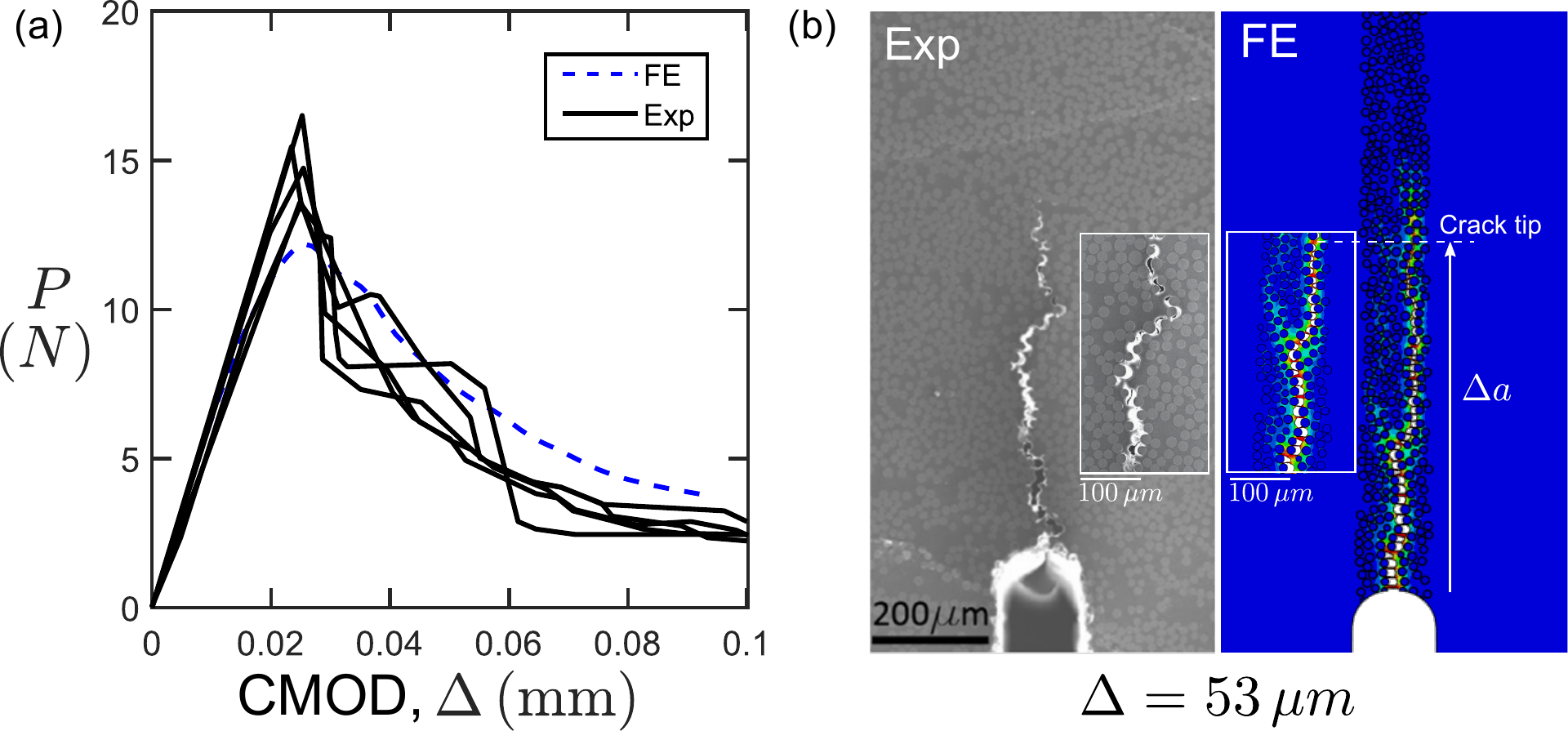}
    \caption{Experimental verification: measured \cite{Canal2012} and predicted (a) load-CMOD curve, and (b) transverse crack propagation in fibre-reinforced composites.}
    \label{fig:ExptValidation}
\end{figure}

We shall first validate model predictions using the single-edge notched three-point bending test depicted in Fig. \ref{Fig.2}. The numerical and experimental results of load versus CMOD response and crack trajectory are summarised in Fig. \ref{fig:ExptValidation}. As shown in Fig. \ref{fig:ExptValidation}a, the computational model accurately predicts the measured load-CMOD displacement curve \cite{Canal2012}, including both elastic and softening regimes. The maximum load is roughly 10\% lower than the measured peak load, which is within the scatter observed in the experiments. The model is also able to capture the microscopic fracture process, see Fig. \ref{fig:ExptValidation}b. A good correlation is found with the observations from the scanning electron micrographs. Fibre-matrix debonding and matrix cracking are the main contributions to the fracture process and a continuous crack path is eventually formed by their coalescence. In close agreement with experimental observations, the numerical results show a large number of matrix ligaments behind the crack. One should note that the stiffness of these matrix ligaments is degraded to almost zero when the phase field damage variable is large, $\phi \approx 1$; in other words, these damaged matrix ligaments do not contribute to crack bridging. 

\subsubsection{2D Grain bridging effect} 

\begin{figure}[htbp]
    \centering
    \includegraphics[width=0.9\textwidth]{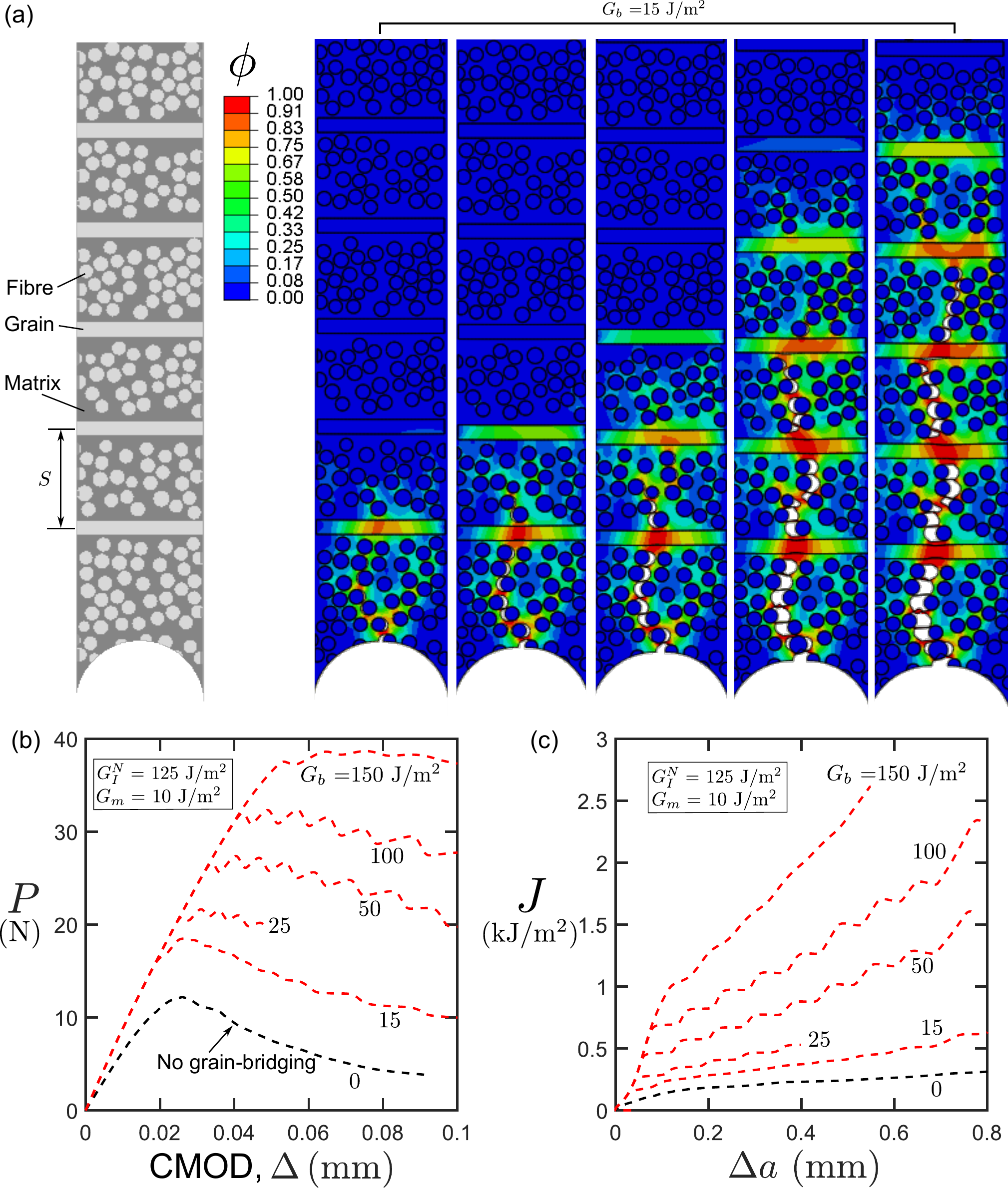}
    \caption{The role of grain bridging on: (a) the crack trajectory, (b) the load-CMOD response, and (c) the fracture resistance curves (R-curves).}
    \label{fig:GrainBridging}
\end{figure}

In most fibre-reinforced composite materials, fibres toughen and strengthen the polymer at the same time. After a crack propagates through the fibre, the fibre elongates and is pulled out from the matrix. These processes correspond to fibre elastic fracture, fibre-matrix debonding and fibre pull-out, which can contribute to the toughening of the composite. To explore the possibility of designing suitable bridging microstrutures, we replace part of the microstructure in the baseline model with bridged fibres of rectangular prismatic shape, henceforth referred to as grains, as shown in Fig. \ref{fig:GrainBridging}a. This shape, not currently used in fibre-reinforced composites, is inspired by biocomposites such as nacre. The interval between the bridged grains equals $S=$ 0.1 mm and the same material properties are assumed for the regular tubular fibres and the rectangular grains. The critical energy release rate of glass fibre $G_b$ is varied between 15 J/m\textsuperscript{2} and 150 J/m\textsuperscript{2}. It should be emphasised that in this work grain bridging is still small scale fibre bridging as the cracked region is small relative to the sample dimensions. In larger scale fibre bridging, the characteristic bridging length is normally a few millimetres in length, which is much larger than the close tip mechanisms studied in this work. In addition, pull-out events of fractured fibres are not considered.  

The predicted load-CMOD curves and R-curve responses are plotted in Fig. \ref{fig:GrainBridging}b and Fig. \ref{fig:GrainBridging}c, with $J$ being computed as described in the introduction Section (small-scale bridging). Compared to the baseline model without grain bridging, bridged fibres significantly improve the modulus, maximum load and R-curve behaviour with increasing fibre fracture toughness $G_b$. The zig-zag patterns in the load-CMOD curves are indicative of the fracture sequence of bridging fibres. When the fibre is elongated, the load increases until fibre breakage occurs. 

\subsubsection{2D Brick-and-mortar microstructure} 

\begin{figure}[H]
    \centering
    \includegraphics[width=0.9\textwidth]{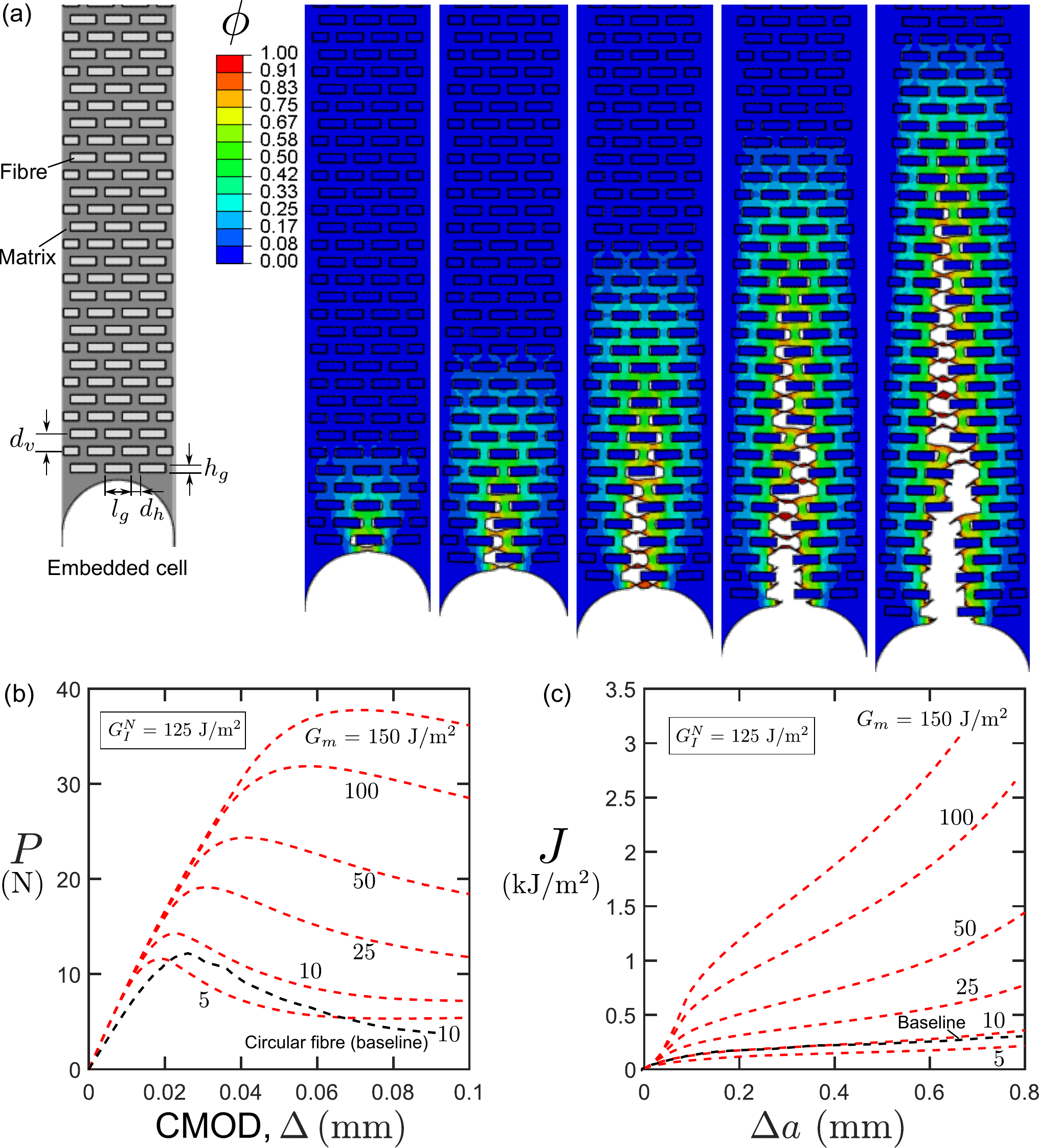}
    \caption{The role of brick-and-mortar microstructure on: (a) the crack trajectory, (b) the load-CMOD response, and (c) the fracture resistance R-curves.}
    \label{fig:BrickMortar}
\end{figure}

Biological materials can provide new sources of inspiration to overcome brittleness. For example, nacre from mollusc shells is a highly regular three-dimensional ``brick-and-mortar'' assembly of microscopic mineral tablets bonded by biopolymers. This structure possesses a fracture toughness significantly larger than its constituents due to toughening mechanisms such as crack deflection and crack branching. Here, we want to explore the effect of the matrix toughness on the macroscopic behaviour of composites based on the brick-and-mortar microstructure. This microstructure is essentially one type of aligned short fibre reinforced composites. The detailed microstructure is shown in Fig. \ref{fig:BrickMortar}a,  where short glass fibres of length $l_g=$ 0.03 mm and width $h_g=$ 0.01 mm fill the epoxy matrix. The fibre volume fraction is equal to $f_g=$ 36.1\% and the distance between adjacent short fibres equals $d_v=0.02$ mm in the vertical direction and $d_h=0.01$ mm in the horizontal direction.

The sequence of crack propagation is assembled in Fig. \ref{fig:BrickMortar}a. A diffused and zig-zag crack path with crack branching and deflection is clearly observed. The predicted microscale fracture consists of fibre-matrix debonding and matrix cracking. Note that the highly distorted matrix ligaments with close-to-zero stiffness have been removed. Again, the bridge zone length $l_b$ is much smaller than the sample dimensions $L$ and $h$; the present brick-and-mortar bridging study is an SSB problem. The predicted load-CMOD and R-curve responses are illustrated in Fig. \ref{fig:BrickMortar}b and Fig. \ref{fig:BrickMortar}c. Again, $J$ is here computed using Eq. (\ref{eq:ssbJ}). The mechanical behaviour and fracture toughness are both enhanced with increasing matrix fracture toughness. The R-curve of the baseline model of circular fibres is comparable to that obtained with the brick-and-mortar structure, assuming the same material properties but a lower fibre volume fraction for the latter. This suggests that adopting a bio-inspired microstructure is a cost-effective method to improve the fracture toughness \cite{Mencattelli2020,Jia2019}. 

\subsection{Three-dimensional, large-scale bridging analyses} 
\label{Sec:LSBResults}

The 2D plane strain simulations presented so far are computationally efficient. However, they can neither represent the circular shape and random distribution of fibres nor capture the stress field around the fibre edge \cite{zhang20203d}. Here, we present the first 3D model of fibre-reinforced composites capable of capturing fibre bridging behaviour.  

\begin{figure}[H]
    \centering
    \includegraphics[width=1.0\textwidth]{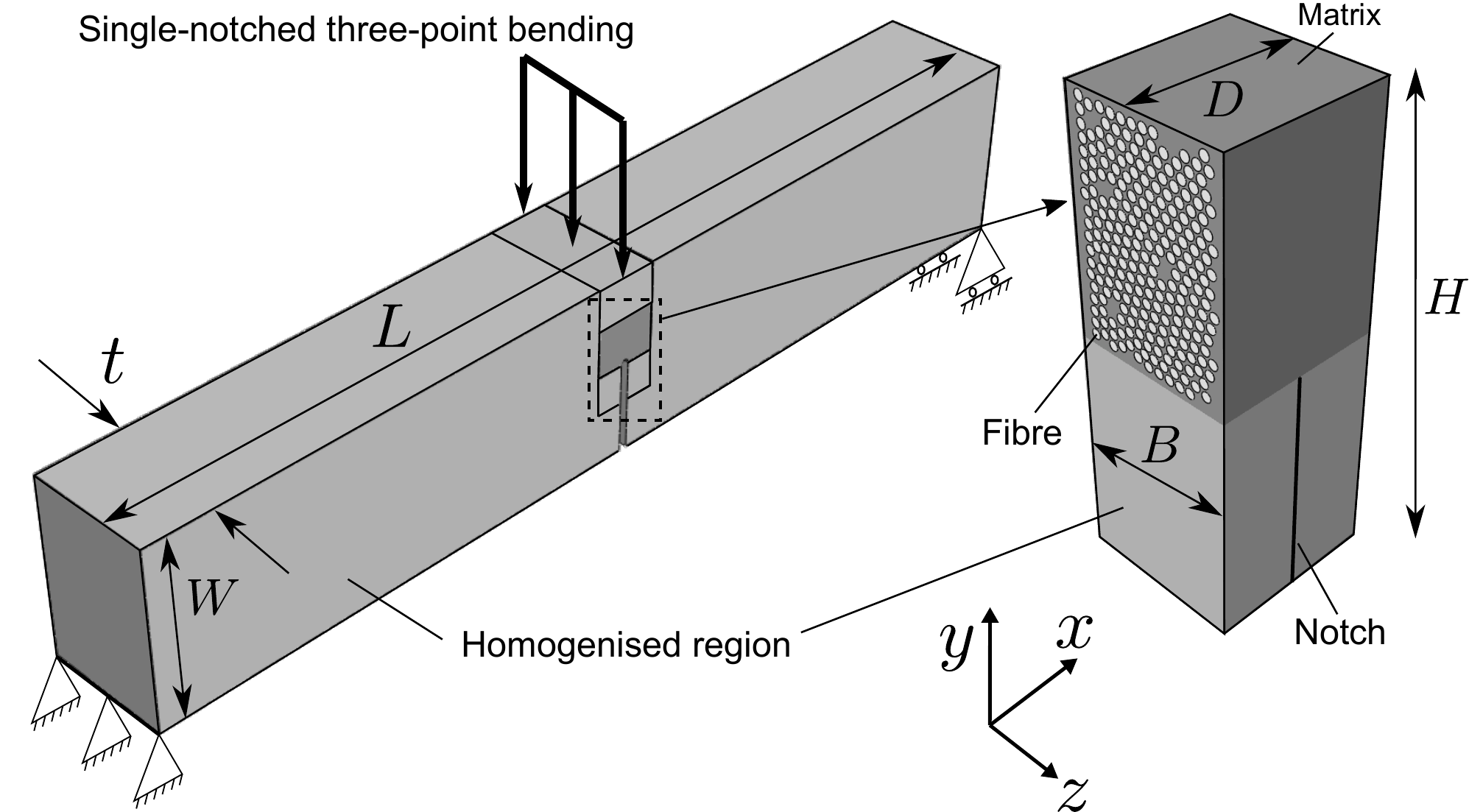}
    \caption{3D large-scale bridging: single-notched three-point bending setup and detail of the embedded cell resolving the composite microstructure.}
    \label{fig:3Dsketch}
\end{figure}

The 3D boundary value problem under consideration is a notched beam undergoing three-point bending, see Fig. \ref{fig:3Dsketch}. The notched beam has the same dimensions as the 2D model in Fig. \ref{Fig.2}, with a thickness of 2 mm. Similar to Section \ref{Sec:SSBResults}, a 3D embedded cell is used to represent the fibre-reinforced composite ahead of the crack tip, while the rest of the region is homogenised as an anisotropic solid. All the fibres are aligned along the $x$ axis, perpendicular to the crack propagation direction, which takes place along the $y$ axis. The dimensions of the representative embedded cell are as follows: thickness $B=$ 0.5 mm, depth $D=$ 0.52 mm and height $H=$ 1.4 mm. A fibre diameter of $d=$ 30 $\si{\micro\metre}$ is assumed and the glass fibre volume fraction ranges from $f_g=$ 13\% to $f_g=$ 48\%. Eight-node brick, hexahedral elements with full integration are used. The element size in the fracture process zone is approximately equal to 5 $\si{\micro\metre}$ and increases up to 20 $\si{\micro\metre}$ at the outer edge of the embedded cell. The total number of elements for the 3D simulations ranges from around 300,000 to 600,000, depending on the fibre volume fraction assumed. This results in a number of degrees-of-freedom (DOFs) between approximately 1.8M and 3.6M. Simulation times varied from 36 to 72 hours, using 32 cores in a single node (Intel Xeon Gold 6126 2.60 GHz, QMUL HPC cluster). Phase field simulations of up to 60M DOFs have been reported for composites fracture using explicit schemes \cite{zhang2021phase,Zhang2021b}. As before, the phase field (PF) fracture model is used to predict cracking in the fibres and the matrix while a cohesive zone model (CZM) is adopted to capture fibre-matrix debonding. The mechanical properties of the fibre, the matrix and the fibre-matrix interface used in the previous section are adopted as baseline material constants. The initial crack length is denoted as $a_0$. We define the current crack tip $a_t$ as the position of the furthest damaged element ($\phi \geq 0.95$), which indicates the crack extension $\Delta a=a_t-a_0$. The end-opening position of the bridging zone $a_e$ is defined as the distance between the crack mouth and the closest unbroken fibre behind the current crack tip ($\phi \leq 0.95$). The bridging zone length is then calculated as $l_b=a_t-a_e$.

\begin{figure}[H]
    \centering
    \includegraphics[width=1.0\textwidth]{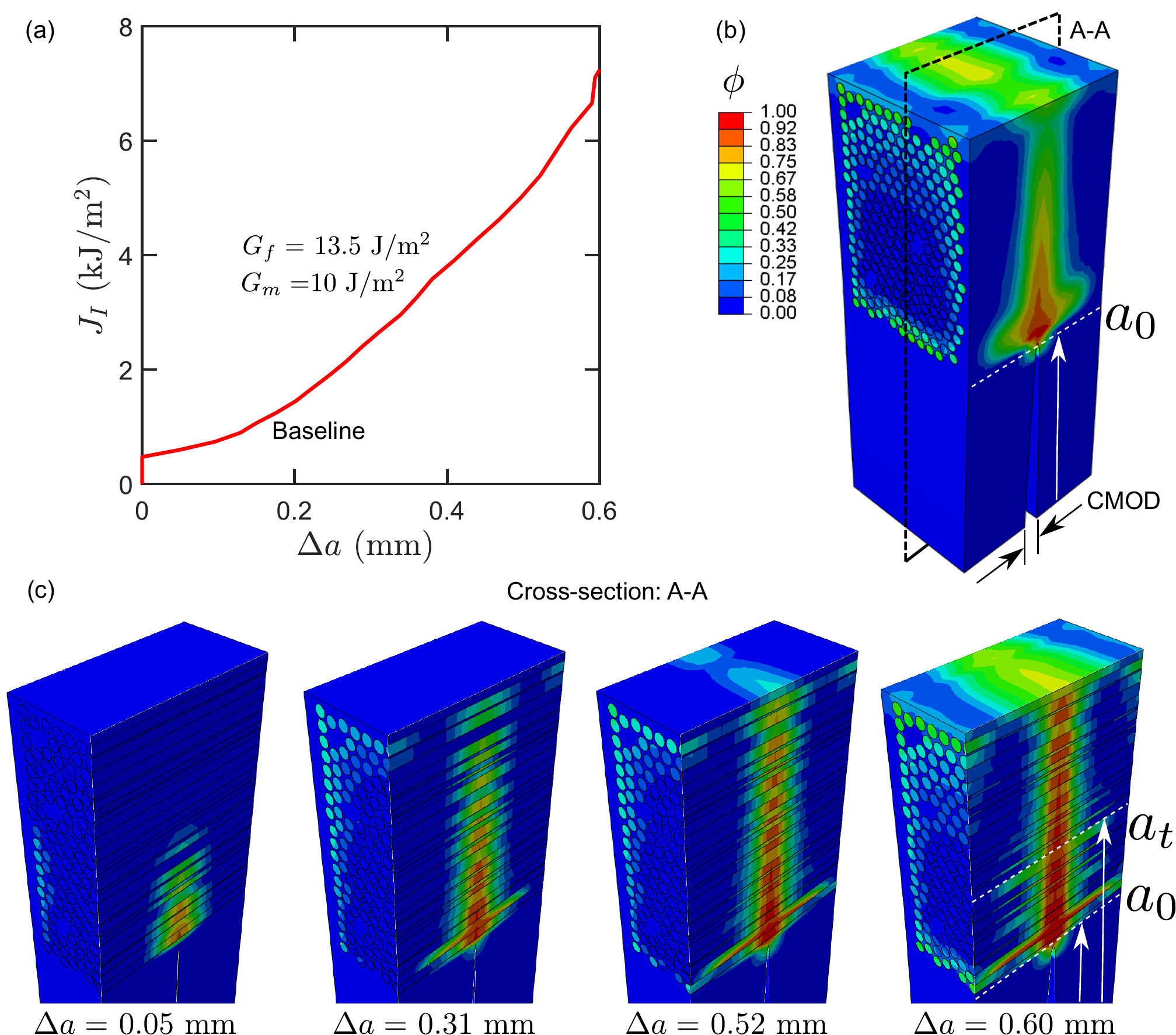}
    \caption{3D large-scale bridging: (a) the role of fibre bridging on the R-curve of composite materials, where $G_f=$ 13.5 J/m\textsuperscript{2}; (b) the phase field damage contour within the embedded cell; and (c) the cross-sectional view of the phase field damage at different crack extensions.}
    \label{Fig.7}
\end{figure}

The predicted crack growth resistance curve (R-curve) is plotted in Fig. \ref{Fig.7}a. In these large-scale bridging analyses, $J$ is estimated using Eq. (\ref{eq:Bridging_law}). The calculated $J$ integral increases significantly with increasing crack extension $\Delta a$, and a substantial degree of subcritical crack growth is observed. Specifically, the magnitude of $J$ increases from 0.47 kJ/m\textsuperscript{2} to 7.1 kJ/m\textsuperscript{2} due to fibre bridging effects. Fig. \ref{Fig.7}b shows an overview of the phase field damage contours in the embedded cell. Cross-sectional snapshots of the phase field damage at different crack extensions are shown in Fig. \ref{Fig.7}c. The main failure mechanisms are matrix cracking and fibre breakage. Fibre-matrix debonding is negligible, therefore very limited fibre pull-out can be observed. Damage initiates by matrix cracking near the notch tip, then fibres start picking up tension and eventually fail. A continuous crack path is developed by the coalescence of matrix cracking and fibre breakage, while the undamaged fibres bridge the crack. A significant spread of damage due to fibre bridging is observed, indicating large energy dissipation via matrix cracking and fibre breakage. Once the results from the baseline model have been established, we proceed to investigate the role of the fibre toughness $G_f$, the fibre volume fraction $f_g$, and the fibre-matrix interface toughness $G_I^N$. The sample dimensions are kept fixed and thus the investigation of phenomena such as the in-situ thickness effect is left for future work \cite{arteiro2014micro,hu2021phase}.


\subsubsection{The effect of fibre fracture toughness}

\begin{figure}[H]
    \centering
    \includegraphics[width=1.0\textwidth]{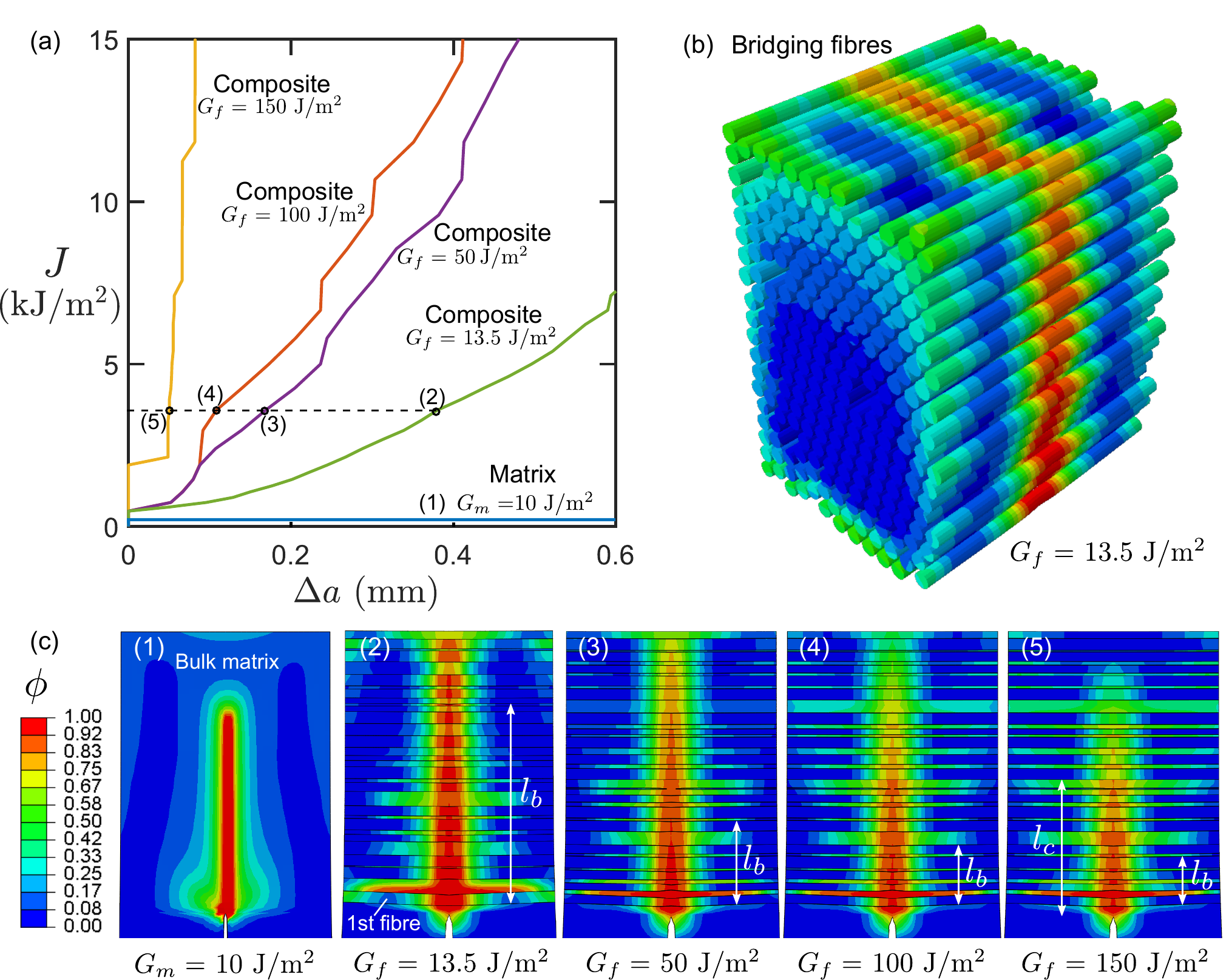}
    \caption{3D large-scale bridging - effect of fibre toughness: (a) the role of fibre bridging on crack growth resistance (R-curves); (b) phase field damage contours in the bridging fibres for $G_f=13.5$ J/m\textsuperscript{2}; and (c) phase field damage contours for the bulk matrix and fibre-reinforced composites for $J=4$ kJ/m\textsuperscript{2}, with the fibre toughness ranging from $G_f=13.5$ J/m\textsuperscript{2} to $G_f=150$ J/m\textsuperscript{2}.}
    \label{fig:fibreToughness}
\end{figure}

The magnitude of the fibre toughness $G_f$ is varied to explore its effect on crack growth resistance. The resulting R-curves are shown in Fig. \ref{fig:fibreToughness}a. The result obtained for the pure epoxy, with $J$ reaching a plateau when equal to $G_m$, showcase the accuracy of the $J$-integral calculations. For the composite material, in agreement with expectations, the energy dissipated during crack growth is greatly enhanced if $G_f$ is increased. The phase field contour within the bridging fibres is illustrated in Fig. \ref{fig:fibreToughness}b, where damage is shown to be mainly located in the middle of the fibre. A comparison of phase field damage contours for the same $J$ magnitude and various fibre toughness values is given in Fig.  \ref{fig:fibreToughness}c. Unsurprisingly, the bulk matrix without any reinforcement shows the longest crack length, whilst the case where the matrix is reinforced with the toughest fibre ($G_f=$ 150 J/m\textsuperscript{2}) exhibits the smallest extension of damage. The bridging zone length $l_b$ is highlighted in all cases. The toughening gains resulting from large-scale fibre bridging, as captured by the 3D model, appear to be much more significant than those resulting from other microstructural mechanisms such as those explored in Section \ref{Sec:SSBResults}, where cracking takes place under small scale bridging conditions and only a few fibres hold the crack faces. In fact, our 3D results reveal that continuous fibre bridging is shown to elevate the energy released at rupture from 10 J/m\textsuperscript{2} to more than 15 kJ/m\textsuperscript{2} - a difference of three orders of magnitude. The fracture toughness of composites can be further enhanced with suitable combinations of material constituents and microstructure, in good agreement with the fracture toughness values reported for intralaminar composites, which range from 1 to 634 kJ/m\textsuperscript{2} \cite{Tan2016,Laffan2012,Marin2016}. We emphasise that these findings are relevant to the conditions relevant to our 3D analyses - composite materials with continuous fibres undergoing longitudinal tensile loading and providing a significant bridging contribution (unlike transverse cracking conditions where matrix cracking and interface debonding are dominant).

\subsubsection{The effect of fibre volume fraction}

\begin{figure}[H]
    \centering
    \includegraphics[width=1.0\textwidth]{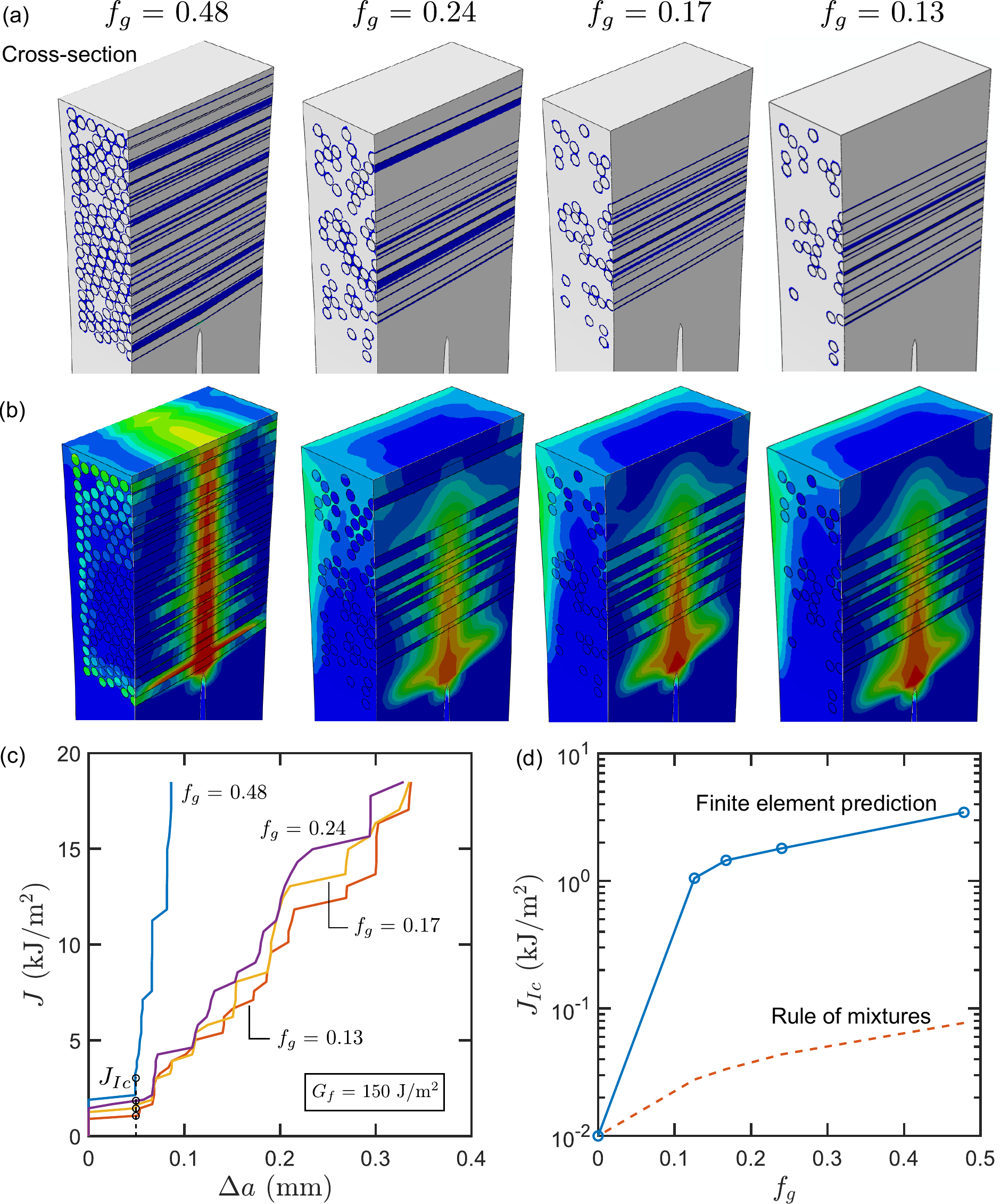}
    \caption{3D large-scale bridging - effect of fibre volume fraction: (a) cross-sectional view of the 3D composite embedded cells; (b) cross-sectional view of the phase field damage contours; (c) R-curves for fibre-reinforced composites with $G_f=150$ J/m\textsuperscript{2} and varying $f_g$; and (d) initiation fracture toughnesses ($J$ at $\Delta a=0.05$ mm) predicted, together with the estimations from the rule of mixtures.}
    \label{fig:Volume fraction}
\end{figure}

We shall now investigate the influence of the fibre volume fraction on fibre bridging toughening. To do so, the fibre volume fraction $f_g$ is varied from 13 \% to 48 \%, as depicted in Fig. \ref{fig:Volume fraction}a. The fracture toughness $J_{Ic}$ in the absence of fibre bridging effects can be estimated using the rule of mixtures, as calculated by,   
\begin{equation}\label{eq:ROM}
    J_{Ic}=f_g J_{gc}+ f_m J_{mc} \, ,
\end{equation} 

\noindent where $f_{m}$ is the volume fraction of matrix, and $J_{gc}$ and $J_{mc}$ are the initiation fracture toughnesses of glass fibre and matrix, respectively. However, if matrix-fibre interaction mechanisms are accounted for, a much higher energy release rate is expected due to a variety of energy dissipation mechanisms, including fibre bridging, fibre-matrix debonding, fibre breakage and matrix cracking. These mechanisms can be captured by our model, as showcased by the phase field damage contours shown in Fig. \ref{fig:Volume fraction}b. The crack growth resistance curves (R-curves) obtained for composites with different fibre volume fractions $f_g$ are shown in Fig. \ref{fig:Volume fraction}c. In agreement with expectations, the $J$-integral increases with increasing fibre volume fraction. We define an initiation toughness $J_{Ic}$ as the value of $J$ at $\Delta a= 0.05$ mm and compute the sensitivity of $J_{Ic}$ to the fibre volume fraction. The results are shown in Fig. \ref{fig:Volume fraction}d, together with the estimations obtained using the rule of mixtures. Differences are significant; by resolving explicitly the fibre bridging mechanisms we can quantify their toughening effect, which leads to $J_{Ic}$ values well in excess of those predicted by the rule of mixtures.

\subsubsection{The effect of fibre-matrix interface toughness}

\begin{figure}[H]
    \centering
    \includegraphics[width=1.0\textwidth]{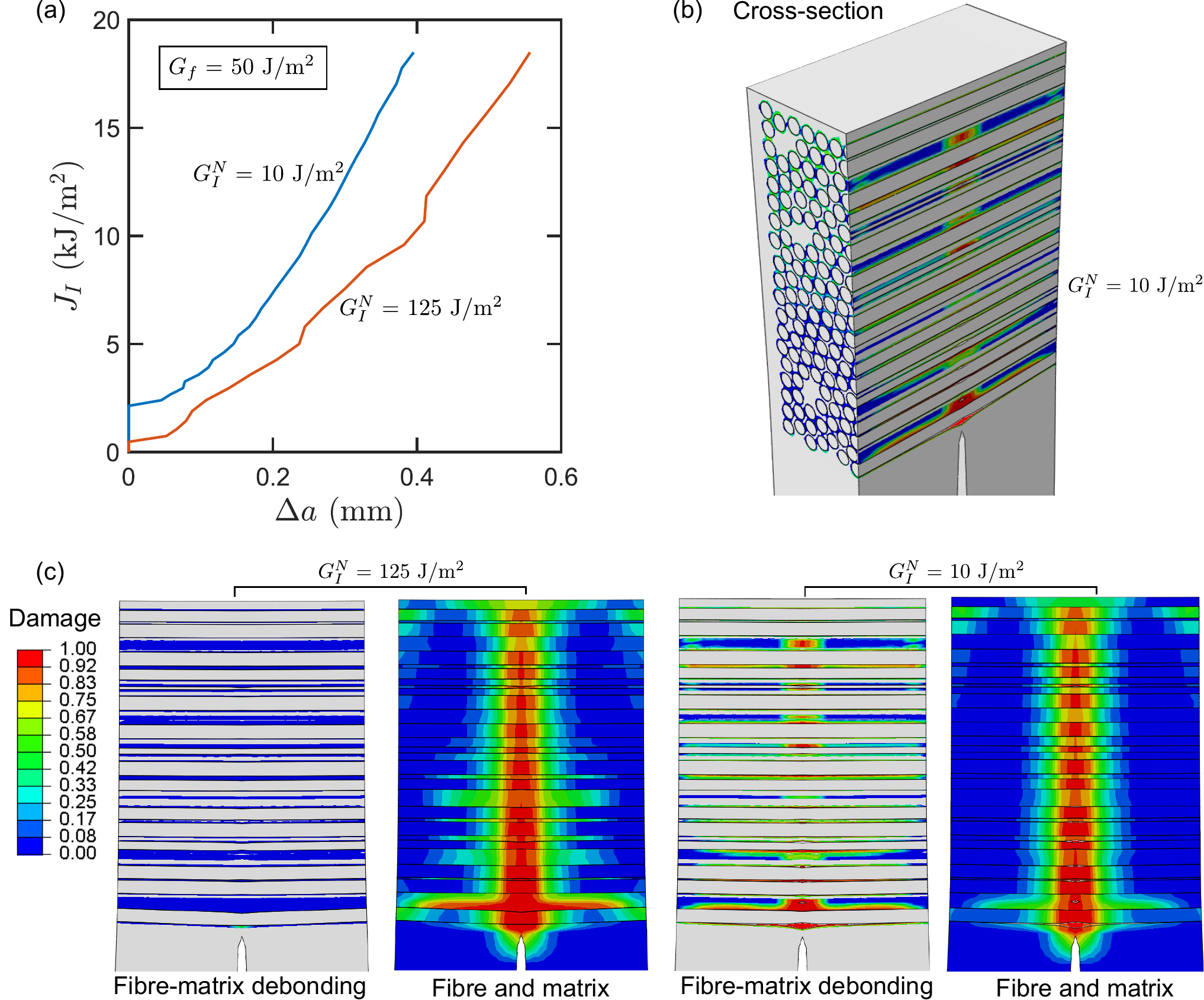}
    \caption{3D large-scale bridging - effect of fibre-matrix interface toughness: (a) R-curves for composite materials with $G_I^N=$ 10 J/m\textsuperscript{2} and $G_I^N=$ 125 J/m\textsuperscript{2}; (b) cross-sectional view of cohesive zone damage contours showing interfacial debonding; and (c) in-plane cross-sectional view of the cohesive zone (left) and phase field (right) damage contours for $G_I^N=$ 10 J/m\textsuperscript{2} and $G_I^N=$ 125 J/m\textsuperscript{2}.}
    \label{Fig.9}
\end{figure}

Finally, the role of the interface toughness is investigated. A weak interface between the matrix and the reinforcing material is beneficial to crack bridging. This strategy has been widely used to toughen brittle materials such as ceramic composites \cite{Xia2004,Jiang2018a}. We quantify the effect of interface toughness $G_I^N$ on the crack growth resistance of fibre-reinforced composites by varying its magnitude from 10 J/m\textsuperscript{2} to 125 J/m\textsuperscript{2}. The toughening resulting from the use of a weak interface only becomes evident when the fracture toughness of the fibre is smaller than the interface fracture toughness ($G_f<G_I^N$). For the case $G_f=$ 50 J/m\textsuperscript{2}, the crack growth resistance curves shown in Fig. \ref{Fig.9}a reveal that a weak interface between the matrix and the fibres augments both the toughness at initiation and the energy released with crack extension. As shown in Fig. \ref{Fig.9}b, a significant amount of fibre-matrix debonding is observed. This leads to redistribution of stress around the fibres and delayed breakage of the fibres, thereby increasing the overall energy dissipation. The damage contours are given in Fig. \ref{Fig.9}c for $G_I^N$=10 J/m\textsuperscript{2} and $G_I^N$=125 J/m\textsuperscript{2}, as predicted by cohesive zone (left, debonding) and phase field (right, fibre and matrix cracking) models. The composite with $G_I^N=$ 125 J/m\textsuperscript{2} exhibits negligible fibre-matrix debonding, with most of the damage being observed in the fibres and the matrix. In the case of $G_I^N=$ 10 J/m\textsuperscript{2}, fibre and matrix cracking is also observed but this is accompanied by numerous fibre-matrix debonding events. This presence of multiple, concurrent failure mechanisms contributes to increasing the crack growth resistance.

\section{Conclusions}
\label{Sec:ConcludingRemarks}

We have proposed a novel computational framework capable of modelling the influence of matrix-fibre microstructures on the crack growth resistance of composites. This is achieved by combining phase field fracture, to capture matrix and fibre cracking, with a cohesive zone model, to simulate matrix-fibre debonding. Fibre-matrix interaction mechanisms, such as fibre-bridging, are explicitly resolved in both 2D and 3D for the first time. Several boundary value problems with various microstructures are simulated to showcase the capabilities of the framework and gain physical insight into the roles of constituent properties and matrix-fibre interactions on macroscopic fracture resistance.

Firstly, the model is validated against data from single-edge notched beam bending experiments. Our predictions are in very good agreement with the experimental results, in terms of both the load-displacement response predicted and the crack trajectory. The role of different microstructures on crack growth resistance is subsequently explored. We show that the use of bridged fibres of rectangular prismatic shape (``grains'') is a very efficient strategy to enhance fracture resistance as a significant amount of energy is dissipated via fibre-matrix debonding and fibre breakage. The design of brick-and-mortar microstructures also shows a noticeable elevation in the fracture toughness, comparable to the continuous fibre reinforcement with a much smaller fibre volume fraction, due to a significant amount of crack deflection and crack branching events.

The simulation of 3D boundary value problems enables gaining a unique insight into fibre toughening mechanisms. Energy dissipating mechanisms such as fibre bridging, fibre-matrix debonding, fibre breakage and matrix cracking can notably elevate toughness predictions. We quantify the role of the fibre fracture toughness, the fibre volume fraction and the fibre-matrix interface toughness. A significant increase in crack growth resistance is observed when weakening the interface, due to increased fibre-matrix debonding, and when increasing the fibre volume fraction and toughness. Toughness values predicted shortly after the onset of crack growth turn out to be several orders of magnitude larger than those predicted using the rule of mixtures. 

The computational framework presented provides a powerful virtual tool to investigate the role of the microstructure and material properties on the fracture behaviour of composite materials and structures of arbitrary geometries and dimensions. This will promote a more efficient design paradigm for improving the fracture toughness of damage-tolerant or energy-absorbing composites.

\section{Acknowledgements}
\label{Sec:Acknowledgeoffunding}

W. Tan acknowledges financial support from the EPSRC [grant EP/V049259/1] and the European Commission Graphene Flagship Core Project 3 (GrapheneCore3) under [grant No. 881603] and the access to QMUL High Performance Computing services. E. Mart\'{\i}nez-Pa\~neda acknowledges financial support from the EPSRC [grant EP/V009680/1] and from UKRI's Future Leaders Fellowship programme [grant MR/V024124/1].



\bibliographystyle{elsarticle-num}

\bibliography{library_2021_12_06}


\end{document}